\documentclass[twocolumn]{aastex63}
\usepackage{dcolumn}
\usepackage{amsmath}
\received{2021 March 31}
\revised{2021 June 25}
\accepted{2021 June 25}
\submitjournal{ApJ}

\begin{document}
\title{Exploring the 100 au Scale Structure of the Protobinary System NGC 2264 CMM3 with ALMA}

\author{Yoshiki Shibayama}
\affiliation{Department of Physics, The University of Tokyo, 7-3-1 Hongo, Bunkyo-ku, Tokyo, 113-0033,Japan}
\author[0000-0002-9668-3592]{Yoshimasa Watanabe}
\affiliation{Materials Science and Engineering, College of Engineering, Shibaura Institute of Technology, 3-7-5 Toyosu, Koto-ku, Tokyo 135-8548, Japan}
\affiliation{Star and Planet Formation Laboratory, RIKEN Cluster for Pioneering Research, Wako, Saitama 351-0198, Japan}
\author{Yoko Oya}
\affiliation{Department of Physics, The University of Tokyo, 7-3-1 Hongo, Bunkyo-ku, Tokyo, 113-0033,Japan}
\affiliation{Research Center for the Early Universe, The University of Tokyo, 7-3-1, Hongo, Bunkyo-ku, Tokyo 113-0033, Japan}
\author{Nami Sakai}
\affiliation{Star and Planet Formation Laboratory, RIKEN Cluster for Pioneering Research, Wako, Saitama 351-0198, Japan}
\author{Ana L\'{o}pez-Sepulcre}
\affiliation{Universit\'{e} Grenoble Alpes, CNRS, IPAG, F-38000 Grenoble, France}
\affiliation{IRAM, 300 rue de la piscine, F-38406 Saint-Martin d’H\'{e}res, France}
\author{Sheng-Yuan Liu}
\affiliation{Institute of Astronomy and Astrophysics, Academia Sinica, 11F of ASMAB, AS/NTU No.1, Sec. 4, Roosevelt Rd, Taipei 10617, Taiwan}
\author{Yu-Nung Su}
\affiliation{Institute of Astronomy and Astrophysics, Academia Sinica, 11F of ASMAB, AS/NTU No.1, Sec. 4, Roosevelt Rd, Taipei 10617, Taiwan}
\author{Yichen Zhang}
\affiliation{Star and Planet Formation Laboratory, RIKEN Cluster for Pioneering Research, Wako, Saitama 351-0198, Japan}
\author{Takeshi Sakai}
\affiliation{Graduate School of Informatics and Engineering, The University of Electro-Communications, Chofu, Tokyo 182-8585, Japan}
\author{Tomoya Hirota}
\affiliation{National Astronomical Observatory of Japan, Mitaka, Tokyo 181-8588, Japan}
\affiliation{Department of Astronomical Sciences, SOKENDAI (The Graduate University for Advanced Studies), Mitaka, Tokyo 181-8588, Japan\vspace*{80pt}}
\author{Satoshi Yamamoto}
\affiliation{Department of Physics, The University of Tokyo, 7-3-1 Hongo, Bunkyo-ku, Tokyo, 113-0033,Japan}
\affiliation{Research Center for the Early Universe, The University of Tokyo, 7-3-1, Hongo, Bunkyo-ku, Tokyo 113-0033, Japan}
\correspondingauthor{Yoshimasa Watanabe}
\NewPageAfterKeywords

\begin{abstract}
We have observed the young protostellar system NGC 2264 CMM3 in the 1.3~mm and 2.0~mm bands at a resolution of about 0\farcs1 (70 au) with ALMA.
The structures of two distinct components, CMM3A and CMM3B, are resolved in the continuum images of both bands.
CMM3A has an elliptical structure extending along the direction almost perpendicular to the known outflow, while CMM3B reveals a round shape.
We have fitted two 2D-Gaussian components to the elliptical structure of CMM3A and CMM3B, and have separated the disk and envelope components for each source.
The spectral index $\alpha$ between 2.0~mm and 0.8~mm is derived to be 2.4-2.7 and 2.4-2.6 for CMM3A and CMM3B, respectively, indicating the optically thick dust emission and/or the grain growth.
A velocity gradient in the disk/envelope direction is detected for CMM3A in the CH$_3$CN, CH$_3$OH, and $^{13}$CH$_3$OH lines detected in the 1.3~mm band, which can be interpreted as the rotation of the disk/envelope system.
From this result, the protostellar mass of CMM3A is roughly evaluated to be $0.1- 0.5$~$M_\sun$ by assuming Keplerian rotation.
The mass accretion rate is thus estimated to be $5\times10^{-5}$ - 4 $\times$ $10^{-3}$ $M_\sun$ yr$^{-1}$, which is higher than typical mass accretion rate of low-mass protostars.
The OCS emission line shows a velocity gradient in both outflow direction and disk/envelope direction.
A hint of outflow rotation is found, and the specific angular momentum of the outflow is estimated to be comparable to that of the disk.
These results provide us with novel information on the initial stage of a binary/multiple system.
\end{abstract}


\section{Introduction}
It is well known that a significant fraction of stars are born in binary or multiple systems \citep[e.g.,][]{Duchene_2013}: the fraction is about a half for the solar-type stars \citep{Raghavan_2010}.
Hence, formation of binary systems is a central issue for star-formation study.
Thanks to developments of radio interferometry, more and more protostellar sources are now found to be binary or multiple systems.
For instance, survey observations of protostars in Perseus and Orion were intensively conducted with ALMA and VLA \citep[e.g.,][]{Tobin_2016, Tobin_2018, Tobin_2020,Maury_2019}, and many binary and multiple protostellar systems have been identified in these observations.
For instance, the multiplicity fraction is reported to be $0.57\pm 0.09$ for the Class 0 protostars in Perseus \citep{Tobin_2016}.  \citet{Maury_2019} reported the multiplicity fraction of $< 0.57 \pm 0.10$ based on the observation of 16 Class 0 protostars in various regions.
Studies on gas kinematics of binary systems have also been conducted by observing molecular emission lines \citep[e.g.,][]{Tobin_2019, Maureira_2020, Oya_2020, Takakuwa_2020}.
To elucidate the formation process of the binary or multiple system, it is still of fundamental importance to reveal its early phase for several prototypical objects under various environmental conditions.
To start with, we here report the case of the very young protostellar source in the high mass star forming region NGC~2264.
\par
NGC 2264 is an active star-forming region near the Solar system \citep[$719\pm 16$~pc:][]{Maiz_2019}.
It belongs to Monoceros OB1 association consisting of hundreds of near infrared sources \citep{Lada_2003}.
The brightest infrared source is IRS1 \citep{Allen_1972} whose mass is reported to be $\sim$9.5 $M_\sun$ \citep{Thompson_1998}.
Around IRS1, 12 dense cores, called CMM1 - CMM12, are distributed within 4$\arcmin$ (1 pc) area, among which CMM3 is the most massive core \citep[40 $M_\sun$:][]{Peretto_2006}.
\citet{Thompson_2000} suggested that CMM3 will form a single high-mass star, or a few intermediate-mass stars.
\citet{Maury_2009} reported that CMM3 will form a single protostar that will evolve into a main sequence star with a mass of 8 $M_\sun$ by using the theoretical evolutionary track.
\citet{Saruwatari_2011} detected an outflow associated with CMM3 whose dynamical timescale is as short as 140-2000 yr.
Since CMM3 is deeply embedded in the parent core, it is thought to be in an early stage of high-mass star formation.
However, the high spatial resolution observation with ALMA revealed that this picture forming a single high-mass star is too simplified.
\citet{Watanabe_2017} conducted sub-arcsecond resolution observations ($\sim$0\farcs3) with ALMA and found that this object is a complex system mainly consisting of CMM3A and CMM3B.
Furthermore, they are surrounded by several faint continuum sources (CMM3C-H).
Therefore, CMM3 is now recognized as a good target to study the early phase of the intermediate-mass binary formation in a cluster-forming environment.
\par
\citet{Watanabe_2017} presented another interesting feature of this source, a significant difference of the spectral appearance between CMM3A and CMM3B.
Very rich molecular lines of complex organic molecules such as CH$_3$OH, HCOOCH$_3$, and (CH$_3$)$_2$O are detected in CMM3A, indicating a hot core/hot corino associated with CMM3A.
On the other hand, a very sparse spectrum is observed for CMM3B.
\citet{Watanabe_2017} proposed the following possibilities of the origin of such a striking difference: (1) chemical difference due to different evolutionary stages of the two protostars, (2) the effect of the dust opacity, and (3) difference of the masses of the two protostars.
A similar case showing the different spectral appearance between the binary components were reported for NGC 1333 IRAS4A \citep{Lopez_2017}.
For this source, it has been reported that the dust opacity causes the difference \citep{Sahu_2019, De_Simone_2020}.
To investigate the NGC 2264 CMM3 case, observations of the dust emission at multiple frequencies are thus awaited.
\par
As described above, NGC 2264 CMM3 is an interesting testbed for exploring physics and chemistry of binary formation in a cluster-formation environment.
For evaluation of the protostellar mass and characterization of the disk/envelope structure for each component, we have observed the CMM3 region with ALMA at an angular resolution of about 0\farcs1, which is three times higher than that of the preceding study.
We have analyzed the distribution of the dust continuum emission at 1.3 mm and 2.0~mm to extract the disk component.
We have also investigated the kinematics of CMM3A by using molecular lines to evaluate its protostellar mass.
This observation will provide us with a clue to discriminating the above possibilities for the different spectral appearance between the two components raised by \citet{Watanabe_2017}.

\section{Observation and Data Reduction}

\begin{deluxetable*}{lcrrcc}[t]
\tablecaption{Emission lines analyzed in this study}
\tablehead{
\colhead{Emission Line} & \colhead{Rest. Freq.\tablenotemark{\rm a}} & \colhead{E$_l$\tablenotemark{\rm a, b}} & \colhead{S$\mu^2$ \tablenotemark{\rm a}} & \colhead{Beam Size} & \colhead{$\sigma$\tablenotemark{\rm c}}\\
\colhead{} & \colhead{(GHz)} & \colhead{(cm$^{-1}$)} & \colhead{(Debye$^2$)} & \colhead{} & \colhead{(Jy beam$^{-1}$)}
}
\startdata
Continuum (1.3 mm)                              &               241&               &          &0\farcs104 $\times$ 0\farcs073 (P.A. = $-$42.9$^\circ$)&1.3 $\times$ 10$^{-4}$\\
Continuum (2.0~mm)                              &               155&               &          &0\farcs069 $\times$ 0\farcs060 (P.A. = $-$38.7$^\circ$)&3.8 $\times$ 10$^{-5}$\\
CH$_3$CN ($14_4-13_4$)                 &257.4481282&135.2681&  395.5&0\farcs112 $\times$ 0\farcs100 (P.A. = $-$55.0$^\circ$)&3.8 $\times$ 10$^{-3}$\\
CH$_3$CN ($14_3-13_3$)                 &257.4827919&100.5247&410.93&0\farcs112 $\times$ 0\farcs100 (P.A. = $-$55.0$^\circ$)&3.8 $\times$ 10$^{-3}$\\
OCS ($20-19$)                                          &243.2180364&77.0793&   10.23&0\farcs119 $\times$ 0\farcs106 (P.A. = $-$53.9$^\circ$)&3.1 $\times$ 10$^{-3}$\\
CH$_3$OH ($16_{3,14}-16_{2,15}$ A)&255.2418880&245.4316&  59.41&0\farcs114 $\times$ 0\farcs103 (P.A. = $-$59.2$^\circ$)&3.5 $\times$ 10$^{-3}$\\
$^{13}$CH$_3$OH ($4_{3,2}-4_{2,3}$ A)&255.2037280&    42.018& 2.9516&0\farcs114 $\times$ 0\farcs103 (P.A. = $-$59.2$^\circ$)&3.5 $\times$ 10$^{-3}$\\
$^{13}$CH$_3$OH ($8_{3,6}-8_{2,7}$ A)&255.2656370&    82.948& 7.2912&0\farcs114 $\times$ 0\farcs103 (P.A. = $-$59.2$^\circ$)&3.5 $\times$ 10$^{-3}$\\
\enddata
\tablecomments{For the continuum emission, the center frequency of the spectral window is given.}
\tablenotetext{\rm a}{Taken from CDMS \citep{Endres_2016}.}
\tablenotetext{\rm b}{Lower state energy.}
\tablenotetext{\rm c}{Root-mean-square (rms) noise level.}
\end{deluxetable*}

The observation in Band 6 (1.3 mm) was carried out on 2019 August 9 with 39 antennas.
The baseline length on the ground ranged from 43.3 to 5893.6 m.
Six spectral windows with a band width of 117 MHz each and a frequency resolution of 0.12 MHz (velocity resolution of 0.14 km s$^{-1}$ at 255 GHz) were employed for molecular line observations, while one spectral window with a bandwidth of 1.875 GHz and a frequency resolution of 0.98 MHz (velocity resolution of 1.2 km s$^{-1}$ at 255 GHz) was mainly used for the continuum observation.
The field center is ($\alpha_{\rm ICRS}$, $\delta_{\rm ICRS}$) = (06$\fh$41$\fm$12$\fs$2800, 09$\fdg$29$\farcm$12$\farcs$000).
The total on-source time was 31.62 min.
The largest recoverable angular scale is 1\farcs3 (930 au).
J0725$-$0054 was used for the bandpass calibration and the flux calibration, while J0643+0857 for the phase calibration.
The molecular lines used in this study are listed in Table~1.
\par
The continuum observation in Band 4 (2.0~mm) was carried out on 2019 July 16 with 46 antennas.
The baseline length on the ground ranged from 92.1 to 8547.6 m.
Four spectral windows with a bandwidth of 938 MHz and a frequency resolution of 0.49 MHz (0.95 km s$^{-1}$ at 155 GHz) were employed.
The field center is the same as that in the band 6 observation.
The total on-source time was 38.42 min.
The largest recoverable angular scale was 1\farcs2 (860 au).
J0725$-$0054 was used for the bandpass calibration and the flux calibration, while J0643+0857 for the phase calibration.
\par
Data reduction was carried out with the Common Astronomy Software Applications package (CASA)\footnote{\url{https://casa.nrao.edu/}}.
Self-calibration was performed for phase and amplitude by using the continuum data and was applied to the line data.
Continuum images were obtained by averaging line-free channels, where the Briggs' weighting with a robustness parameter of 0.0 is employed for the CLEAN process.
Six data cubes for the molecular lines listed in Table~1 were obtained after subtracting the continuum component from the visibility data and were CLEANed with automasking and Briggs weighting with a robustness parameter of 0.5.
The synthesized beam sizes are summarized in Table 1.
Velocity resolutions of all the cubes were set to 0.25 km s$^{-1}$.
The root-mean-square (rms) noises of the images are shown in Table 1.
We estimate an uncertainty of the absolute flux scale to be $10$~\%\footnote{ALMA Cycle 7 Technical Handbook \\(\url{https://almascience.nrao.edu/documents-and-tools/cycle7/alma-technical-handbook/})} in the observations with Band~4 and Band~6.

\section{Results and Discussion}

\subsection{Continuum images}
\begin{figure*}[t]
\centering
\includegraphics[clip,width=18cm,bb = 0 0 1050 550, scale = 1.0]{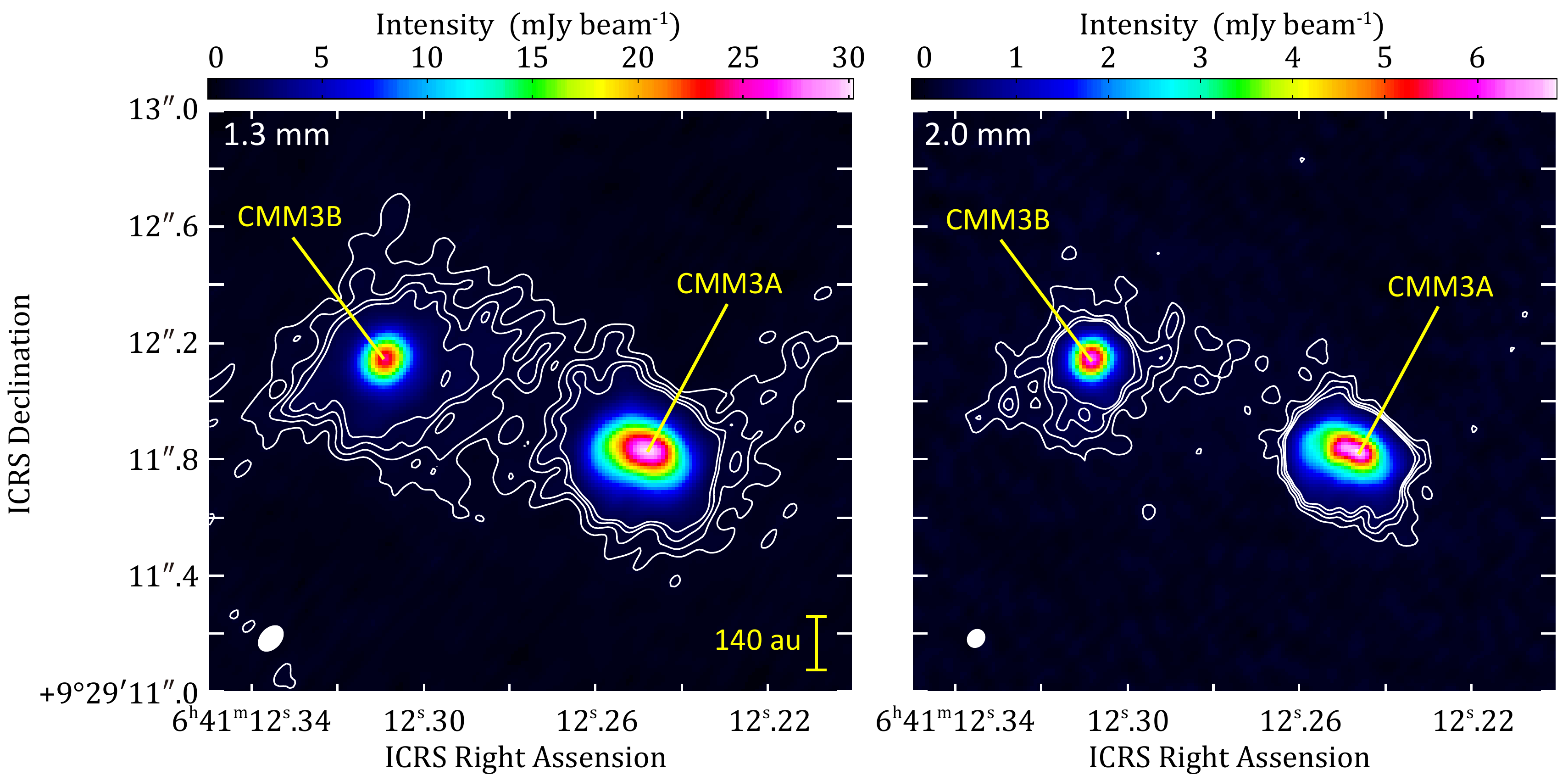}
\caption{The 1.3 mm (Band 6) and 2.0~mm (Band 4) continuum images of CMM3A and CMM3B (Band 6 and Band 4). The white ellipse at the lower left corner of each image represents the beam size. Contours levels are $\sigma$ $\times$ (3, 5, 7, 9). $\sigma$ = 0.13 mJy~beam$^{-1}$ for 1.3 mm and 3.8 $\times$ 10$^{-2}$ mJy~beam$^{-1}$ for 2.0~mm.}
\label{fig01}
\end{figure*}

Figure~\ref{fig01} shows the 1.3 mm (Band 6) and 2.0~mm (Band 4) continuum images.
Two dust continuum peaks, CMM3A and CMM3B, are clearly seen.
The two sources are well separated by 0\farcs95 (680 au), although they look loosely connected with each other by a weak parental envelope, as seen in the 1.3 mm image.
\par
The total fluxes at 1.3 mm of CMM3A and CMM3B over 0\farcs6 $\times$ 0\farcs6 area are 191 $\pm$ 19 mJy and 99 $\pm$ 10 mJy, respectively.
The sum of these two fluxes is smaller than the total 1.3 mm flux of CMM3 (472 mJy) observed with  the 3\farcs18$\times$2\farcs67 beam of SMA even if we take into account of the flux uncertainty of 20~\% in the observation with SMA \citep{Saruwatari_2011}.
Furthermore, it is much smaller than the 1.2~mm flux (1108 mJy) and 1.3~mm flux (1500 mJy) observed with single dish telescopes by \citet{Peretto_2006} and \citet{Thompson_2000}, respectively.
Contrary to these previous studies, we are just looking at much smaller structures with the largest recoverable scale of 1\farcs3.
Hence, the extended component of CMM3 is heavily resolved-out in this observation.
\par
Assuming that the dust continuum emission is optically thin, the total gas mass of CMM3A and CMM3B are evaluated to be 0.11-0.24~$M_\sun$ and 0.05-0.11~$M_\sun$, respectively, from the 1.3~mm data by using the following equation:
\begin{equation}
M_{\rm g} = \frac{S_\nu d^2}{\kappa_\nu B_\nu (T_{\rm d})R_{\rm d}}\hspace{1mm},
\end{equation}
where $M_{\rm g}$ is the total gas mass, $S_\nu$ is the total flux, $d$ is the distance to NGC 2264 \citep[$719 \pm 16$ pc:][]{Maiz_2019}, $\kappa_\nu$ is the frequency-dependent dust mass opacity coefficient, $B_\nu(T_{\rm d})$ is the black-body function of Planck's law at the dust temperature $T_{\rm d}$, and $R_{\rm d}$ is the dust-to-gas mass ratio.
$\kappa_\nu$ and $R_{\rm d}$ are assumed to be 1.3~cm$^2$ g$^{-1}$ at 1.3~mm and 0.01, respectively, as adopted by \citet{Lopez_2017}.
The $\kappa_\nu$ value is originally given in \citet{Ossenkopf_1994}.
$T_{\rm d}$ is assumed to be in the range of 100-200 K.
If the dust continuum emission is optically thick, the total gas mass derived above is regarded as the lower limit.
\par
CMM3A has an elliptical structure extending in the east-west direction.
To define its distribution, we first try to fit a single 2D-Gaussian function to the 2.0~mm continuum image, where the effect of the synthesized beam is taken into account.
However, we find that a fairly large residual is left in the image after subtracting the single 2D-Gaussian model, as shown in \ref{fig03}(f).
A similar residual is seen in the single 2D-Gaussian fit to the 1.3~mm continuum image (See Appendix).
The residuals seem to originate from the excess emission in the southern part.
Then, we conduct double 2D-Gaussian fitting on the image plane, which means the fitting with two 2D-Gaussian components.
Figures \ref{fig03}(b-e) show the result.
The systematic residuals in the single 2D-Gaussian fitting are almost eliminated, as shown in Figure~\ref{fig03}(e) (Also see Appendix for the 1.3~mm data).
The fitted Gaussian components are shown in Figures~\ref{fig03}(c, d) for the 2.0~mm image, while those for the 1.3~mm image are in Appendix.
The fitting parameters are summarized in Table 2. 
CMM3A consists of compact and extended components.
The compact component can be interpreted as the disk structure, while the extended component as the surrounding envelope structure.
Both components extend in a direction almost perpendicular to the outflow of CMM3A, which blows along the P.A. of $-5^\circ$ \citep{Saruwatari_2011, Watanabe_2017}.

\begin{widetext}
\movetabledown=4.0cm
\begin{rotatetable}
\begin{deluxetable*}{cccccccccc}
\tablecaption{The fitting parameters of the double 2D-Gaussian fitting\tablenotemark{\rm a}}
\tablehead{
\colhead{Band} & \colhead{Source} & \colhead{Component} & \colhead{Major Axis\tablenotemark{\rm b}} &
\colhead{Minor Axis\tablenotemark{\rm b}} & \colhead{P.A.\tablenotemark{\rm b}} & \colhead{R.A.} & \colhead{Decl.} & \colhead{Peak Intensity}\\
\colhead{} & \colhead{} & \colhead{} & \colhead{(FWHM)} &
\colhead{(FWHM)} & \colhead{} & \colhead{(ICRS)} & \colhead{(ICRS)} & \colhead{}
}
\startdata
Band 4&CMM3A&Compact&0\farcs164 $\pm$ 0\farcs001&0\farcs059 $\pm$ 0\farcs001&62.8$^\circ$ $\pm$ 0.3$^\circ$&06$\rm ^h$41$\rm ^m$12$\rm ^s$.246&09$^\circ$29$^\prime$11$^{\prime\prime}$.835&3.79 mJy beam$^{-1}$
\\
 & & &($118 \pm 3$ au)&($42.3 \pm 0.9$ au)&&&&\\
          &             &Extended&0\farcs275 $\pm$ 0\farcs001&0\farcs1915 $\pm$ 0\farcs0009&76.5$^\circ$ $\pm$ 0.4$^\circ$&06$\rm ^h$41$\rm ^m$12$\rm ^s$.250&09$^\circ$29$^\prime$11$^{\prime\prime}$.818&3.38 mJy beam$^{-1}$
\\
 & & &($198 \pm 4$ au)&($138 \pm 3$ au)&&&&\\
          &CMM3B&Compact&0\farcs0997 $\pm$ 0\farcs0003&0\farcs0979 $\pm$ 0\farcs0003&87$^\circ$ $\pm$ 6$^\circ$&06$\rm ^h$41$\rm ^m$12$\rm ^s$.308&09$^\circ$29$^\prime$12$^{\prime\prime}$.146&6.04 mJy beam$^{-1}$
\\
 & & &($72 \pm 2$ au)&($70 \pm 2$ au)&&&&\\
          &            &Extended&0\farcs547 $\pm$ 0\farcs006&0\farcs382 $\pm$ 0\farcs004&105.7$^\circ$ $\pm$ 0.7$^\circ$&06$\rm ^h$41$\rm ^m$12$\rm ^s$.308&09$^\circ$29$^\prime$12$^{\prime\prime}$.080&0.37 mJy beam$^{-1}$
\\
 & & &($393 \pm 9$ au)&($276 \pm 6$ au)&&&&\\
Band 6&CMM3A&Compact&0\farcs190 $\pm$ 0\farcs002&0\farcs067 $\pm$ 0\farcs002&62.3$^\circ$ $\pm$ 0.3$^\circ$&06$\rm ^h$41$\rm ^m$12$\rm ^s$.247&09$^\circ$29$^\prime$11$^{\prime\prime}$.837&14.6 mJy beam$^{-1}$
\\
 & & &($137 \pm 3$ au)&($48 \pm 1$ au)&&&&\\
          &            &Extended&0\farcs266 $\pm$ 0\farcs001&0\farcs192 $\pm$ 0\farcs001&68.8$^\circ$ $\pm$ 0.4$^\circ$&06$\rm ^h$41$\rm ^m$12$\rm ^s$.250&09$^\circ$29$^\prime$11$^{\prime\prime}$.815&17.2 mJy beam$^{-1}$
\\
 & & &($191 \pm 4$ au)&($138 \pm 3$ au)&&&&\\
          &CMM3B&Compact&0\farcs1072 $\pm$ 0\farcs0003&0\farcs0994 $\pm$ 0\farcs0004&60.7$^\circ$ $\pm$ 1.4$^\circ$&06$\rm ^h$41$\rm ^m$12$\rm ^s$.309&09$^\circ$29$^\prime$12$^{\prime\prime}$.145&22.4 mJy beam$^{-1}$
\\
 & & &($77 \pm 2$ au)&($71 \pm 2$ au)&&&&\\
          &            &Extended&0\farcs538 $\pm$ 0\farcs005&0\farcs393 $\pm$ 0\farcs004&104.4$^\circ$ $\pm$ 0.7$^\circ$&06$\rm ^h$41$\rm ^m$12$\rm ^s$.306&09$^\circ$29$^\prime$12$^{\prime\prime}$.100&2.1 mJy beam$^{-1}$
\\
 & & &($386 \pm 9$ au)&($283 \pm 6$ au)&&&\\\hline
\enddata
\tablecomments{Errors are standard deviations except for the sizes of major and minor axes in au whose errors are mainly determined by the uncertainty of the distance to NGC 2264 \citep[$719\pm 16$ pc:][]{Maiz_2019}}.
\tablenotetext{\rm a}{The fitting is done on the image plane.}
\tablenotetext{\rm b}{Beam-deconvolved size.}
\end{deluxetable*}
\end{rotatetable}
\end{widetext}
\par
Assuming that the compact component is an inclined thin disk, the inclination angle $i$ with respect to the line of sight (i.e., $i$ = 90$^\circ$ for the edge-on case) is evaluated to be 65$^\circ$ from the deconvolved sizes of the major and minor axes.
This inclination angle is regarded as the lower limit, because the thickness of the disk is ignored.
If the thickness is considered, the inclination angle can be larger to reproduce the sizes of the major and minor axes.
Thus, the inclination of the disk of CMM3A likely has a nearly edge-on configuration.
This result is consistent with the previous observation of the outflow by \citet{Saruwatari_2011}.
We then evaluate the thickness of the inner envelope.
Assuming that the extended component is an inclined thick disk with the same inclination angle as that of the compact component, the thickness $h$ is evaluated to be 70-90 au.
This value is about a third of the size of the major axis (200 au), and hence, the inner envelope seems to have a flattened structure.
\par
After subtracting the fitted double 2D-Gaussian functions from the observed images, a faint residual pattern still remains in the eastern (X) and southwestern(Y) sides of CMM3A in both 2.0~mm and 1.3~mm data, the former of which is shown in Figure~\ref{fig03}(e) (See Appendix for the 1.3~mm data).
These patterns may represent internal structures of CMM3A.
In addition, the two peaks of the residual ($\alpha$ and $\beta$) can be seen in the compact component in the 2.0~mm data (Figure~\ref{fig03}(e)), which are symmetrically located with respect to the peak of the compact Gaussian component.
This means that the compact component has a flat-topped or double-peaked structure.
Such a structure is not seen in Band 6 probably because of the lower angular resolution and/or higher dust opacity.
This structure may be caused by a saturation effect due to high opacity.
Alternatively, it may represent a substructure within the compact component or a possibility of a close binary system with a separation of about 70 au.
\par
The structure of CMM3B was also fitted better with double 2D-Gaussian functions than with a single 2D-Gaussian function (Figures~\ref{fig05} for the 2.0~mm data and Appendix for the 1.3~mm data).
The fitting parameters are summarized in Table 2.
The distribution is composed of the compact and extended components.
In contrast to the CMM3A case, the compact component has a nearly round shape.

\begin{figure*}
\centering
\includegraphics[clip,width=18cm,bb = 0 0 1200 1350, scale = 1.0]{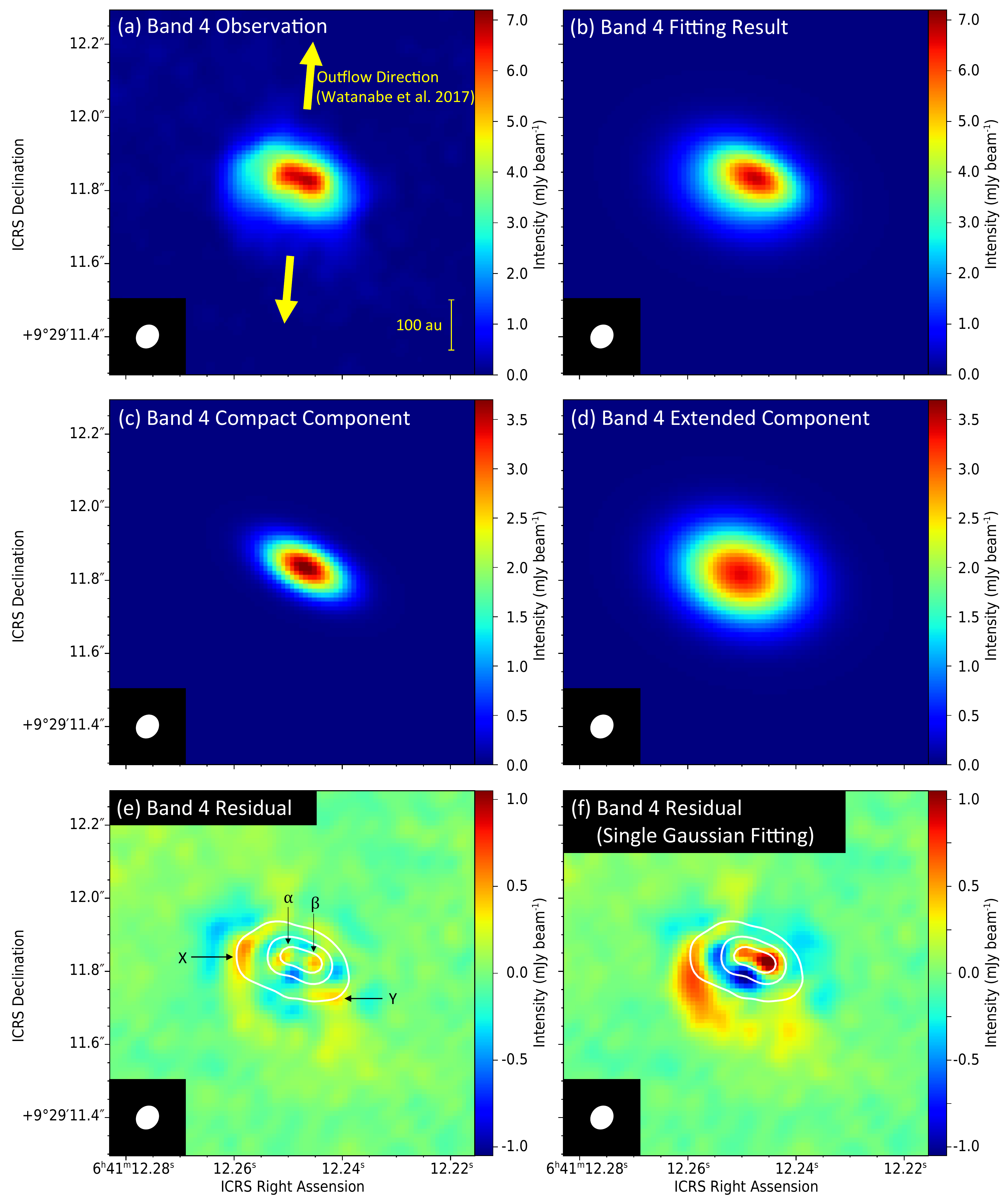}[t]
\caption{(a) The 2.0~mm (Band 4) continuum image of CMM3A.
(b) The result of double 2D-Gaussian fit.
(c), (d) The compact and extended components of the fitted double 2D-Gaussian convolved with the beam.
(e) Residuals of the double 2D-Gaussian fitting.
(f) Residuals of the single 2D-Gaussian fitting for comparison.
The continuum emission at 2.0~mm is displayed in contours with steps of 50$\sigma$ where $\sigma$ is 3.8 $\times$ 10$^{-2}$ mJy~beam$^{-1}$ in (e) and (f).
X and Y denote the position of the systematic residuals also seen in the residuals of double 2D-Gaussian fit for the 1.3~mm continuum (See Appendix).
For $\alpha$ and $\beta$, see text (Section 3.1).}
\label{fig03}
\end{figure*}

\begin{figure*}
\centering
\includegraphics[clip,width=18cm,bb = 0 0 1200 1350, scale = 1.0]{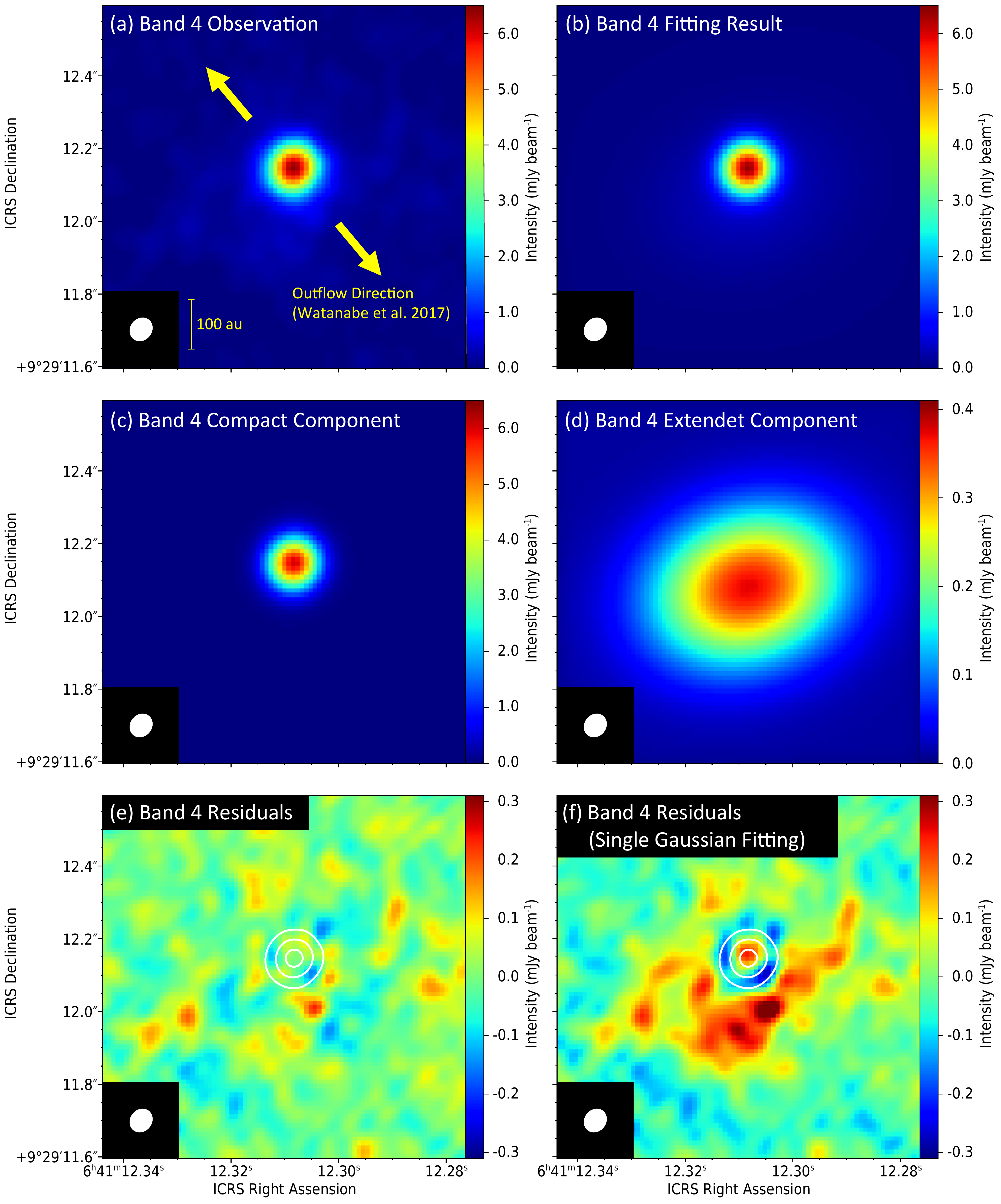}
\caption{(a) The 2.0~mm  (Band 4) continuum image of CMM3B.
(b) The result of double 2D-Gaussian fit.
(c), (d) The compact and extended components of the fitted double 2D-Gaussian convolved with the beam.
Note that the color scale is different between (c) and (d).
(e) Residuals of the double 2D-Gaussian fitting.
(f) Residuals of the single 2D-Gaussian fitting for comparison.
The continuum emission at 2.0~mm is displayed in contours with steps of 50$\sigma$ where $\sigma$ is 3.8 $\times$ 10$^{-2}$ mJy~beam$^{-1}$ in (e) and (f).
}
\label{fig05}
\end{figure*}

\begin{figure*}[t]
\centering
\includegraphics[clip,width=18cm,bb = 0 0 1500 600, scale = 1.0]{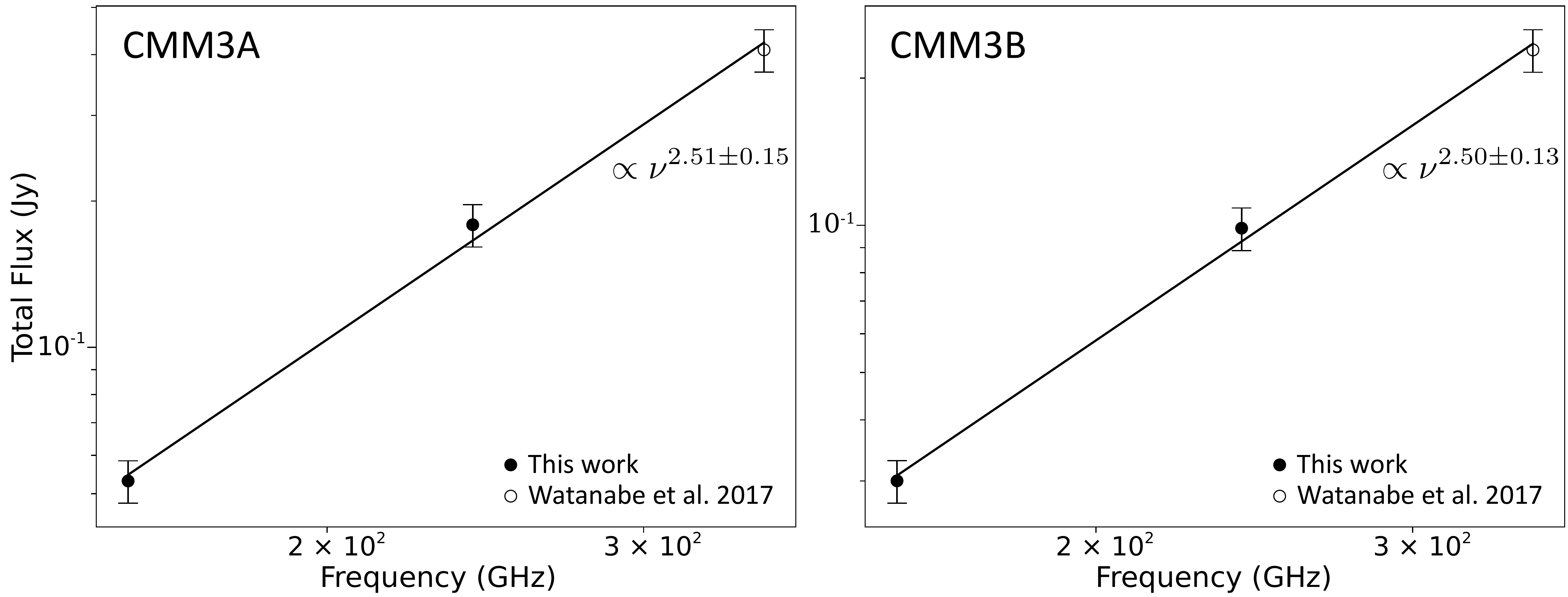}
\caption{The spectral index derived from total flux.
The closed circles represent the results of this work, while the open circle represent the 0.8 mm data taken from \citet{Watanabe_2017}.
The horizontal axis shows the frequency of the continuum images and the longitudinal axis shows the total flux. 
Errors are 10\% of each total flux.
The flux is measured for 0\farcs6 $\times$ 0\farcs6 area for CMM3A and CMM3B.}
\label{fig06}
\end{figure*}

\begin{deluxetable*}{ccccc}[t]
\tablecaption{Spectral indices.}
\tablehead{
\colhead{Source} & \colhead{Component} & \colhead{Spectral Index\tablenotemark{\rm a}} & \colhead{Area\tablenotemark{\rm b}} & \colhead{Used Bands (mm)}
}
\startdata
CMM3A &Total      &2.4-2.7 & 0\farcs6 $\times$ 0\farcs6 & 2.0,1.3,0.8 \\
      &Compact    &2.2-3.1 & 0\farcs6 $\times$ 0\farcs6 & 2.0,1.3\\
      &Extended   &2.3-3.2 & 0\farcs6 $\times$ 0\farcs6 & 2.0,1.3\\
      &Peak       &2.2-2.5 & 0\farcs38 $\times$ 0\farcs31, pa=-35.4$^{\circ}$&2.0,1.3,0.8\\
CMM3B &Total      &2.4-2.6 & 0\farcs6 $\times$ 0\farcs6 & 2.0,1.3,0.8\\
      &Compact    &1.9-2.8 & 0\farcs6 $\times$ 0\farcs6 & 2.0,1.3\\
      &Extended   &2.6-3.5 & 0\farcs6 $\times$ 0\farcs6 & 2.0,1.3\\
      &Peak       &1.9-2.4 & 0\farcs38 $\times$ 0\farcs31, pa=-35.4$^{\circ}$&2.0,1.3,0.8\\
\enddata
\tablenotetext{\rm a}{Ranges are derived from the estimated errors of the flux (10\%).}
\tablenotetext{\rm b}{The area in which the flux is evaluated.}
\end{deluxetable*}

\subsection{Continuum Spectrum}
In this subsection, we investigate the spectral energy distribution (SED) by using the 1.3 mm and 2.0~mm continuum images in combination with the 0.8 mm (350 GHz) image previously observed  with ALMA \citep{Watanabe_2017}.
We use the total fluxes over 0\farcs6 $\times$ 0\farcs6 area for CMM3A and CMM3B to cover their extended structures. 
The two areas for the flux measurement are not overlapped with each other, because the separation between CMM3A and CMM3B is 0\farcs95.
The disk and envelope structures of CMM3A are not distinguished here.
\par
Figure~\ref{fig06} shows the SED plots for CMM3A and CMM3B.
The power-law index ($\alpha$) of the spectral energy distribution ($I(\nu) \propto \nu^\alpha$) derived from total fluxes is 2.4-2.7 for CMM3A and 2.4-2.6 for CMM3B.
These values are rather small and close to the Rayleigh-Jeans limit ($\alpha = 2$).
In interstellar clouds, the $\alpha$ index is mostly in a range of 3.5-4.0 \citep{Li_2001}.
It decreases as dust drains grow in protoplanetary disks \citep[e.g.,][]{Perez_2015}.
Grain growth has recently been reported for disk structures in the low-mass Class 0 and Class I sources \citep[e.g.,][]{Friesen_2018, Gerin_2017}.
Thus, the grain growth could be a possible cause of the small $\alpha$ indices.
Alternatively, the $\alpha$ index can be close to 2, if the dust continuum emission is optically thick.
An effect of the dust scattering may also contribute to the small index \citep{Liu_2019}.
Indeed, the peak brightness temperature of the continuum emission at 1.3 mm is 76-92 K and 62-76 K for CMM3A and CMM3B, respectively.
Considering that CMM3A and CMM3B are expected to be young ($\sim$1000 yr) protostars, we think that the observed spectral indices rather imply the high dust opacities of CMM3A and CMM3B.
We will come back to this point along with results of molecular line observations (Section 3.3).
\par
For further investigation, we evaluate the spectral indices of the compact and extended components by using the 1.3 mm and 2.0~mm total fluxes over 0\farcs6 $\times$ 0\farcs6 area.
Table 3 shows the result, where the fluxes are derived by integrating those of 2D-Gaussian model within the area.
The spectral indices of both components are 2.2-3.2 for CMM3A.
For CMM3B, the spectral index of the compact and extended components are 1.9-2.8 and 2.6-3.5, respectively.
Despite large uncertainties, the results are almost consistent with those derived for the whole structures of CMM3A and CMM3B by using the three bands.
In addition, we also derive the spectral indices for CMM3A and CMM3B by using the fluxes measured with the 0\farcs381 $\times$ 0\farcs308 area (Table~3).
The spectral indices thus obtained are slightly smaller than those derived for the 0\farcs6 $\times$ 0\farcs6 area because of the less contribution of the extended part of the envelope having a larger index.

\subsection{Molecular Distribution}
\begin{figure*}[t]
\centering
\includegraphics[clip,width=17.5cm, bb = 0 0 900 920, scale = 0.9]{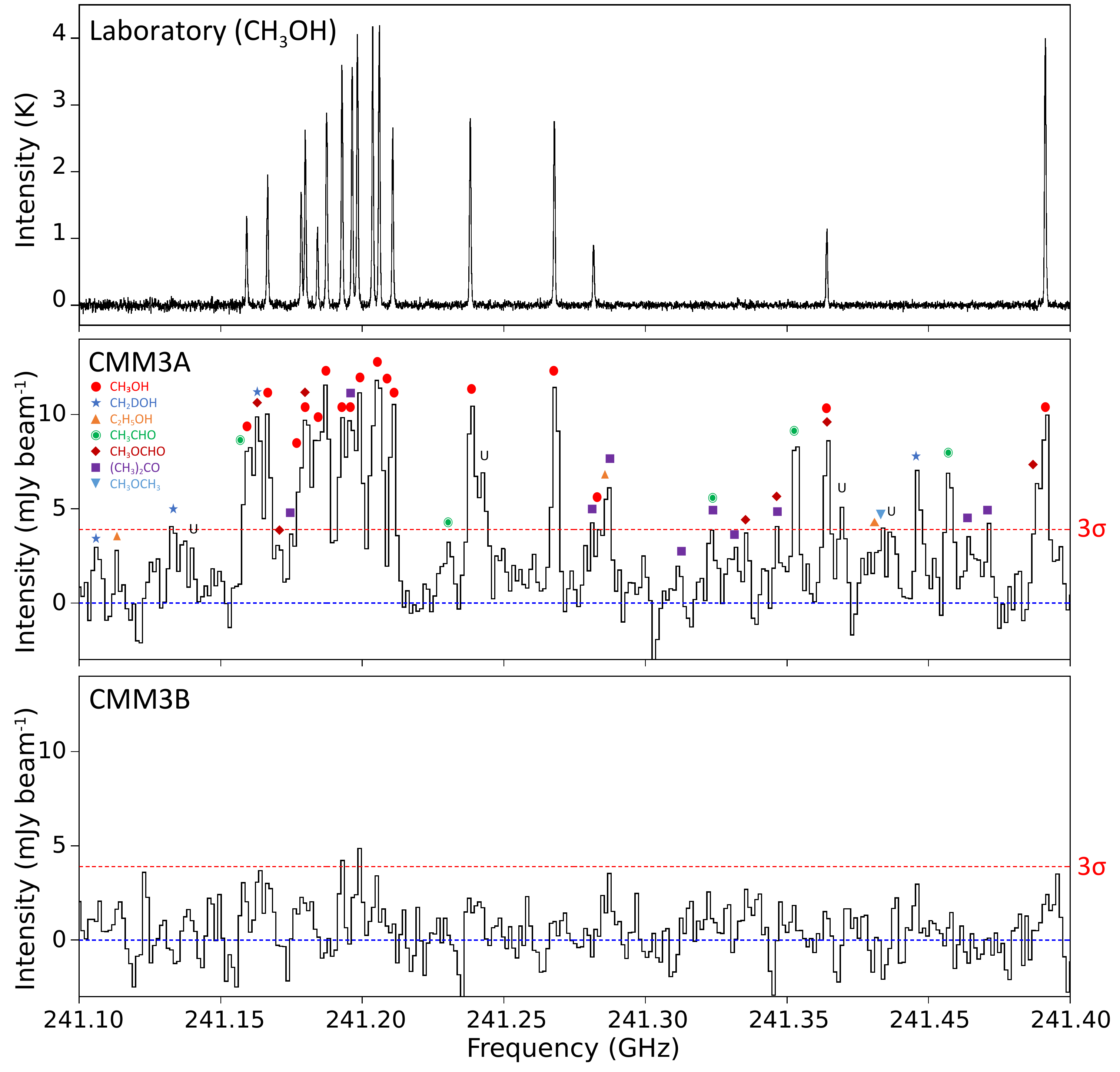}
\caption{(Middle and Bottom) Spectra observed at the continuum peaks of CMM3A and CMM3B.
The red line represents the 3$\sigma$ level ($\sigma$ = 2.0 K).
The frequency resolution is 0.98 MHz.
(Top) CH$_3$OH lines observed at the room temperature in the laboratory \citep{Watanabe_2021}.
Molecular lines are rich in CMM3A and deficient in CMM3B as found by \citet{Watanabe_2017}.}
\label{fig07}
\end{figure*}

\begin{figure*}[t]
\includegraphics[clip,width=18cm,bb = 0 0 1750 900, scale = 1.0]{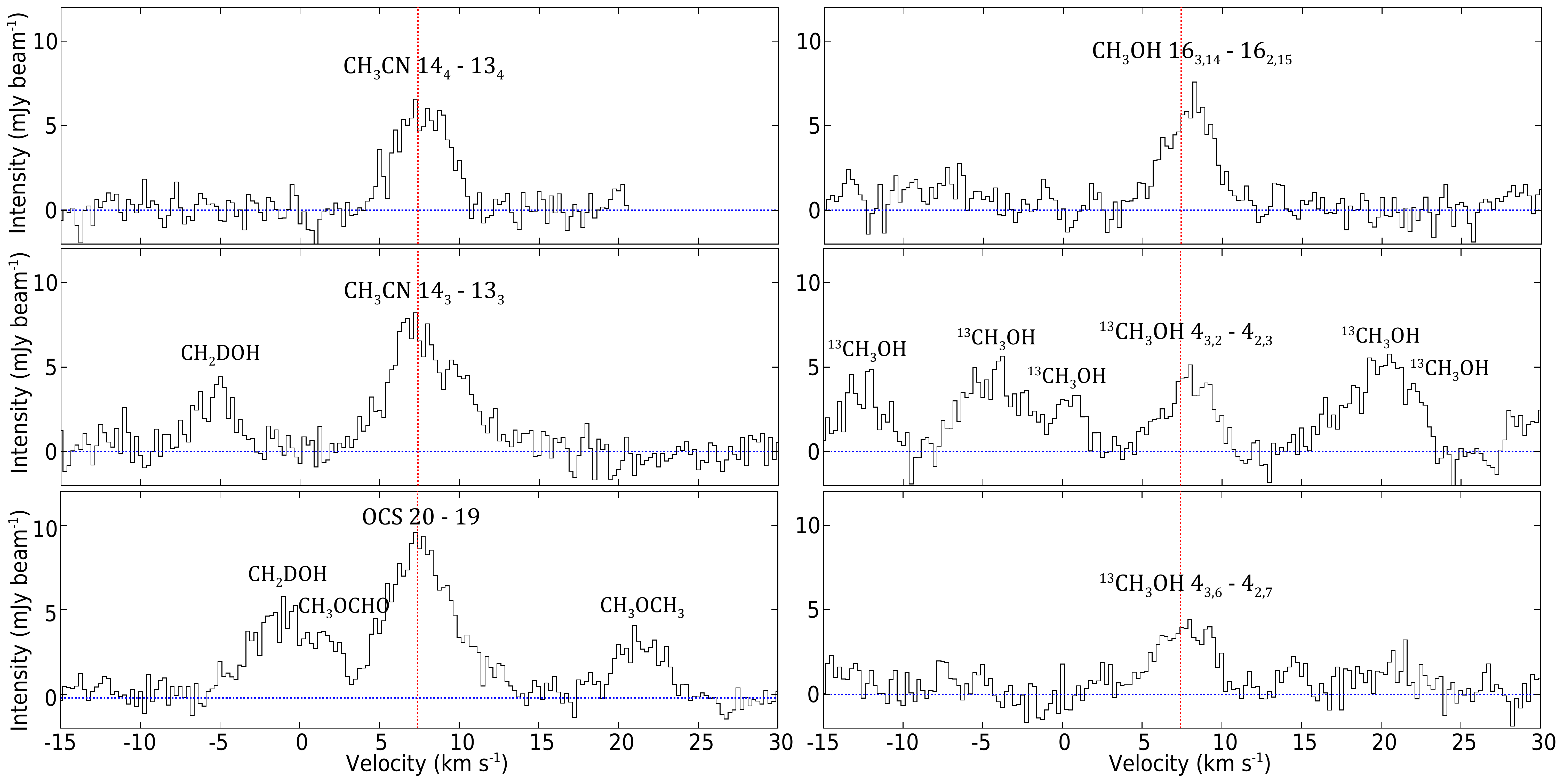}
\caption{Spectral profiles of the molecular lines listed in Table 1, which are observed toward CMM3A.
The spectra are prepared by averaging the 0\farcs6 $\times$ 0\farcs6 area.
The velocity resolution is 0.25 km s$^{-1}$.
The vertical red line represents the systemic velocity of CMM3A \citep[7.4 km s$^{-1}$:][]{Watanabe_2017}.
}
\label{fig08}
\end{figure*}

Figure~\ref{fig07} shows the spectra observed at the continuum peaks of CMM3A and CMM3B in the 1.3 mm band.
The spectra are prepared by using (a part of) the spectral window for the continuum observation with the synthesized beam.
Here, we employ the systemic velocities of 7.4~km~s$^{-1}$ and 7.8~km~s$^{-1}$ for CMM3A and CMM3B, respectively \citep{Watanabe_2017}.
The spectrum of CH$_3$OH observed at room temperature in laboratory is also shown for reference \citep{Watanabe_2021}.
Molecular emission lines in this figure mostly come from the torsionally excited CH$_3$OH ($v_A$ = 1).
The spectrum is rich in CMM3A but sparse in CMM3B, as found in the 0.8 mm band by \citet{Watanabe_2017}.
Because of the congested spectrum of CMM3A, molecular lines are often contaminated by other molecular lines.
To investigate the kinematic structure, we select six uncontaminated molecular lines among the lines observed at a high frequency resolution as listed in Table 1.
Here, line contamination is checked by using the spectral line databases: The Cologne Database for Molecular Spectroscopy \citep[CDMS:][]{Endres_2016} and the Jet Propulsion Laboratory (JPL) catalog \citep{Pickett_1998}.
Figure~\ref{fig08} shows the spectral profiles of the selected lines observed at CMM3A.

\par
\begin{figure*}[t]
\centering
\includegraphics[clip,width=19cm,bb = 0 0 1500 1100, scale = 1.0]{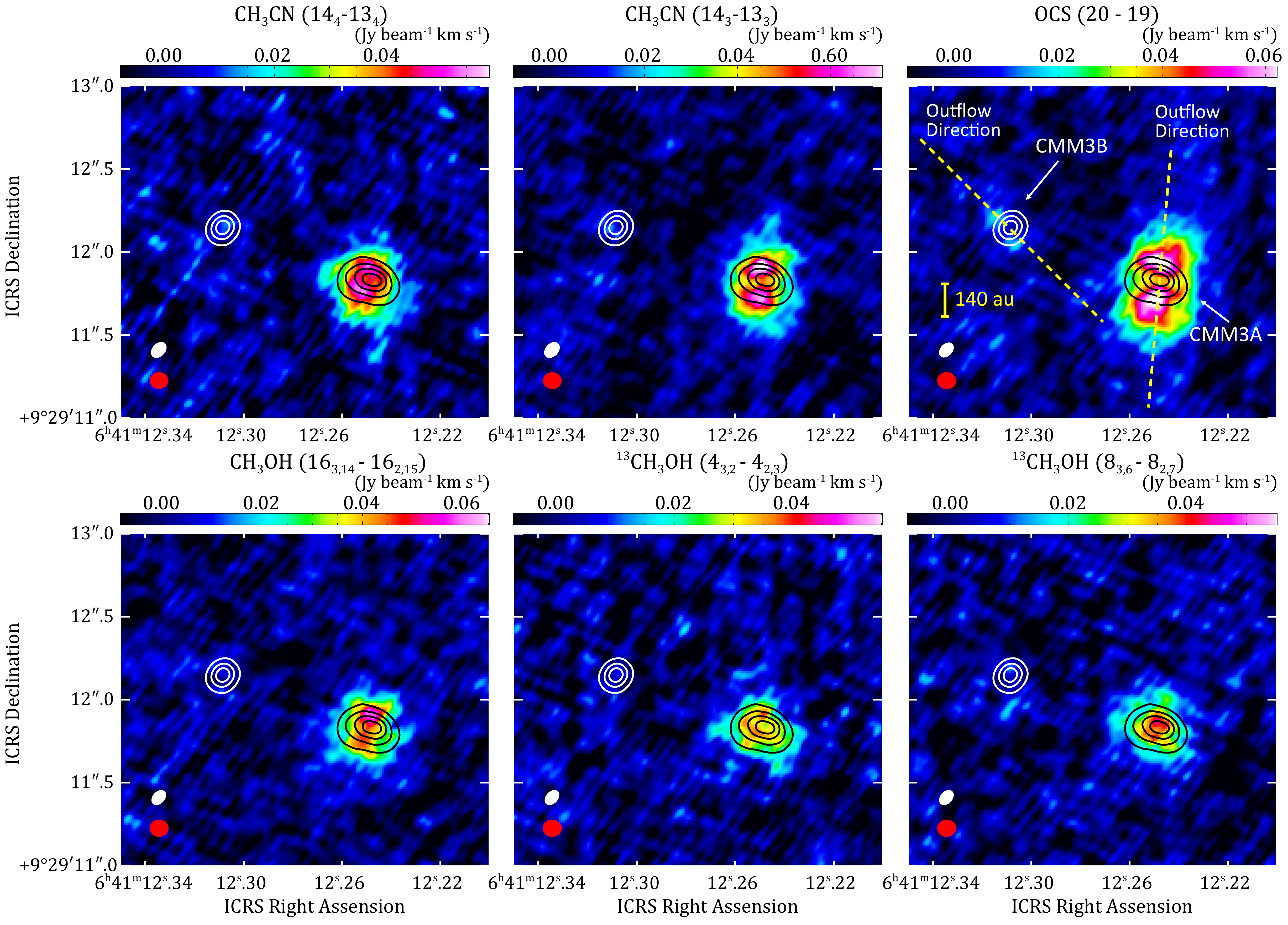}
\caption{The 0th moment maps of the emission lines listed in Table 1.
The velocity range of integration is from 2.5~km~s$^{-1}$ to 12.5~km~s$^{-1}$.
The rms noises are  $4.9 \times 10^{-3}$~Jy~beam$^{-1}$km~s$^{-1}$ for OCS ($20-19$), $5.5 \times 10^{-3}$~Jy~beam$^{-1}$km~s$^{-1}$ for CH$_3$OH ($16_{3,14}-16_{2,15}$), $^{13}$CH$_3$OH ($4_{3,2}-4_{2,3}$), and $^{13}$CH$_3$OH ($8_{3,6}-8_{2,7}$), and $6.0 \times 10^{-3}$~Jy~beam$^{-1}$km~s$^{-1}$ for CH$_3$CN ($14_4-13_4$) and CH$_3$CN ($14_3-13_3$).
The continuum emission at 1.3 mm is displayed in contours with steps of 50$\sigma$ where $\sigma$ is 0.13 mJy beam$^{-1}$.
The white ellipse represent the beam size of the 1.3 mm continuum image.
The red ellipse represents the beam size of each 0th moment map.
The yellow dashed lines indicate outflow directions which are -5$^\circ$ and 50$^\circ$ in P.A., for CMM3A and CMM3B, respectively, based on the results reported by \citet{Watanabe_2017}.}
\label{fig09}
\end{figure*}
\begin{figure*}
\centering
\includegraphics[clip,width=16cm,bb = 0 0 1600 1900, scale = 1.0]{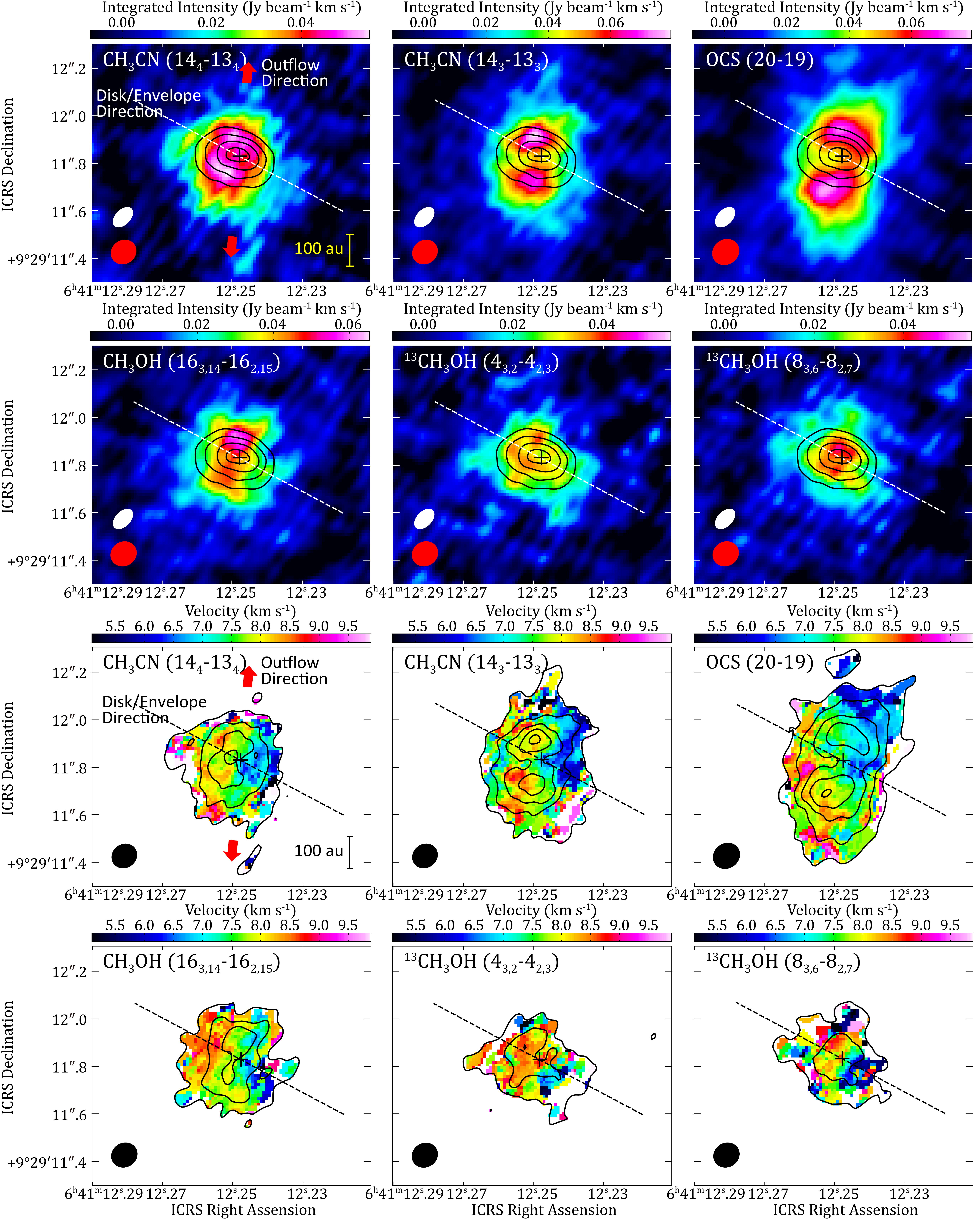}
\caption{The 0th and 1st moment maps of the emission lines listed in Table 1 at CMM3A.
The velocity range of integration is from 2.5~km~s$^{-1}$ to 12.5~km~s$^{-1}$.
The black ellipse represents the beam size.
The black crosses represents the protostellar position of CMM3A.
The continuum emission at 1.3 mm is displayed on the 0th moment maps by contours with steps of 50$\sigma$ where $\sigma$ is 0.13 mJy beam$^{-1}$.
The contours on the 1st moment maps are 3$\sigma$, 6$\sigma$, 9$\sigma$ and 12$\sigma$ of the 0th moment map of each emission line.
The region where the 0th moment is lower than 3$\sigma$ is clipped in the 1st moment maps.
The red arrows indicates outflow directions of CMM3A (-5$^\circ$ in P.A). (See caption of Figure~\ref{fig09}).
The disk/envelope direction is estimated from the 1.3 mm continuum images (P.A. = 62$^\circ$).}
\label{fig10}
\end{figure*}

Figure~\ref{fig09} shows the 0th moment maps (integrated intensity maps) of the six selected lines in the 1.3 mm band.
Figure~\ref{fig10} depicts the 0th moment maps expanded around CMM3A and the corresponding 1st moment maps.
As shown in Figure~\ref{fig09}, emission lines are detected in CMM3A, while no lines except for the faint OCS ($J=20-19$) line are detected in CMM3B with the rms sensitivity of 5 mJy beam$^{-1}$ km s$^{-1}$.
These emission lines detected in CMM3A show a velocity gradient along the east-west direction (redshifted in the eastern side and blueshifted in the western side), as revealed in the 1st moment maps Figure~\ref{fig10}.
For CH$_3$OH and $^{13}$CH$_3$OH, the velocity gradient is almost along the elongated shape of the continuum images, and hence, the velocity gradients are most likely interpreted as the rotation motion of CMM3A.
On the other hand, the gradient for CH$_3$CN and OCS seems to be a combination of the outflow and the rotation of the disk/envelope system, considering that the outflow of CMM3A blows along the P.A. of $-5^\circ$ \citep{Saruwatari_2011, Watanabe_2017}.


\par
It should be noted that the 0th moment maps of CH$_3$CN ($14_4-13_4$  and $14_3-13_3$), OCS ($20-19$), and CH$_3$OH ($16_{3, 14}-16_{2, 15}$ A) show the distributions extending to the north and south directions from the protostar with slight depression along the midplane of the disk/envelope structure where the continuum emission is bright.
In contrast, the 0th moment maps of $^{13}$CH$_3$OH do not clearly show such features.
Hence, the depression along the disk mid-plane seems to originate from the high opacity of dust continuum and molecular line emissions.
The observed intensity is approximately proportional to $\{1-\exp{(-\tau_{\nu})}\}$, where $\tau_{\nu}$ is approximately the sum of the optical depth of the molecular lines and that of the dust.
In the optically thin case ($\tau_{\nu}$ $\ll$ 1), the intensity is proportional to the optical depth, and the observed intensity is the sum of the intensity of the emission line and that of the continuum image.
After subtracting the continuum emission, just a line image is observed.
However, in optically thick case, the observed intensity is saturated at a fixed value and not proportional to the optical depth.
Thus, the observed intensity is not the sum of the intensity of the emission line and that of the continuum image.
Since we obtain data cubes for molecular lines by subtracting the intensity of the continuum image from the observed intensity, the line intensity becomes lower in this case for the region where the intensity of the continuum is high.
This situation is actually reported in HH212, which is known as a "hamburger"-shaped dusty disk: the intensities of the molecular emissions show depression in the midplane of the disk \citep{Codella_2018, Lee_2018}.
\par
Judging from the spectral index derived from the continuum images, the dust opacity of CMM3A would be high in the disk region.
On the basis of the outflow morphology \citep{Saruwatari_2011}, the disk of CMM3A is nearly edge-on (inclination angle $\sim$60$^\circ$).
Therefore, the dust opacity could be particularly high around the midplane of the disk.
This seems to contribute to the depression of the intensity along the disk structure.
The emission lines of CH$_3$CN, OCS, and CH$_3$OH are expected to be optically thick around the mid-plane of the disk/envelope structure, while those of $^{13}$CH$_3$OH are not.
In Figure~\ref{fig07}, the CH$_3$OH emission lines in CMM3A indeed look almost saturated to a certain level of intensity, although the laboratory spectrum shows intensity variation from line to line.
The CH$_3$OH$(16_{3,14}-16_{2,15}\,{\rm A})$ line should be optically thick, as the $^{13}$CH$_3$OH$(8_{3,6}-8_{2,7}\,{\rm A})$ line is observed with the comparable intensity.
The optical depth of the CH$_3$OH line is roughly estimated to be 50 or larger, assuming the LTE condition with the temperature higher than 100~K.
Moreover, the CH$_3$CN $14_3-13_3$ emission is expected to be brighter by a factor of 2 or more than the $14_4-13_4$ emission, because the nuclear spin degeneracy of the former emission is as twice large as that of the latter emission.
Nevertheless, these emission lines have similar intensities in Figure~\ref{fig08}, which indicates high optical depths of these lines.
A similar case for the CH$_3$CN lines is often reported in high-mass star-forming regions \citep[e.g.,][]{Furuya_2011, Ilee_2018}.
Thus, the line optical depth would also contribute to the intensity depression.
\par
Likewise, the very sparse spectral lines toward CMM3B could be interpreted as the effect of the high dust optical depth, as in the case of NGC 1333 IRAS 4A1 \citep{De_Simone_2020}.
Centimeter-wave observations of the continuum and line emissions are essential to confirm this interpretation.

\subsection{Rotation of the CMM3A disk}
\begin{figure*}[t]
\centering
\includegraphics[clip,width=10cm,bb = 0 0 900 950, scale = 1.0]{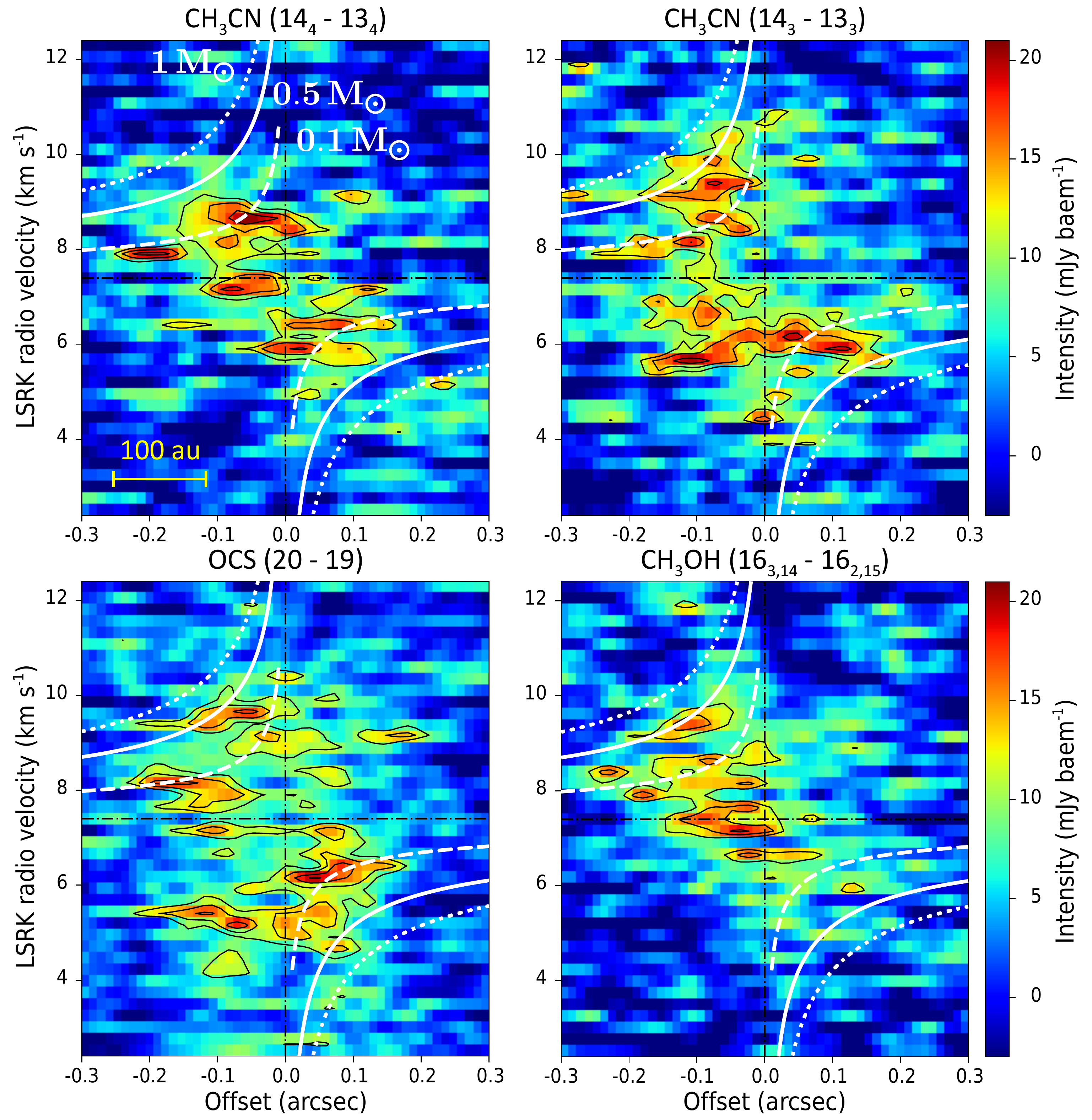}
\caption{The PV diagrams of CH$_3$CN, OCS, and CH$_3$OH.
The contours are $3\sigma$, $4\sigma$, and $5\sigma$, where $1\sigma$ is $3\times 10^{-3}$~Jy~beam$^{-1}$. 
The PV diagrams are made along the disk/envelope direction estimated from the 1.3 mm and 2.0~mm continuum images (P.A. = 62$^\circ$).
The horizontal axis is the distance from the 1.3mm continuum peak, and the vertical axis is the velocity along the line of the sight.
The horizontal line represents the systemic velocity of CMM3A \citep[7.4km s$^{-1}$:][]{Watanabe_2017}.
Dashed, solid, and dotted curves represent the Keplerian rotation with the protostellar mass of 0.1, 0.5, and 1 $M_\sun$, respectively.
The inclination angle is assumed to be 65$^\circ$.
\label{fig11}
}
\end{figure*}

We find a rotation motion of the disk for CMM3A in the CH$_3$CN, OCS, CH$_3$OH, and $^{13}$CH$_3$OH lines (Section 3.3).
To analyze their velocity structures in detail, we prepare the position-velocity (PV) diagrams of CH$_3$CN, OCS, and CH$_3$OH along the disk/envelope direction (Figure~\ref{fig11}).
Although the PV diagrams are patchy due to a poor signal-to-noise ratio, the velocity gradient along the disk/envelope direction is indeed detected for these lines.
It is, however, difficult to distinguish between the Keplerian rotation and the infalling rotating motion conserving the angular momentum \citep[cf.][]{Oya_2016, Oya_2018}.
Then, we roughly estimate the protostellar mass of CMM3A, assuming the Keplerian rotation as:
\begin{eqnarray}
|v| = \sqrt{\frac{GM_{\rm ps}}{r}}\sin{i}.
\end{eqnarray}
Here, $v$ is the velocity along the line of the sight, $G$ is the gravitational constant, $M_{\rm ps}$ is the mass of the protostar, $r$ is the distance from the protostar, and $i$ is the inclination angle of the disk.
The Keplerian velocities for the masses of 0.1 $M_\sun$, 0.5 $M_\sun$ and 1 $M_\sun$ are plotted on Figure~\ref{fig11}, where the inclination angle of the disk is assumed to be 65$^\circ$.
It is difficult to determine the mass precisely because of a poor S/N ratio of the PV diagram.
Nevertheless, the mass of the protostar is roughly constrained to be 0.1-0.5 $M_\odot$ by comparing the rotation curves with the observed PV diagrams.
Since the protostar is deeply embedded in a massive and accreting parent core \citep{Maury_2009}, the relatively low protostellar mass means the infancy of the protostar.
\par
\citet{Saruwatari_2011} found the dynamical timescale of the outflow of CMM3A is 140-2000 yr.
Dividing the estimated protostellar mass of $0.1-0.5$ $M_\sun$ by the dynamical timescale, the average mass accretion rate of CMM3A is estimated to be 5 $\times$ 10$^{-5}$ - 4 $\times$ 10$^{-3}$ $M_\sun$ yr$^{-1}$.
This rate is about 10-100 times higher than the theoretical mass accretion rate of low-mass protostars \citep[10$^{-6}$-10$^{-5}$ $M_\sun$ yr$^{-1}$:][]{Larson_2003}.
According to the numerical simulations of the protostellar evolution including episodic accretion \citep{Machida_2011, Vorobyov_2006}, the mass accretion rate in the earliest evolutionary stage up to 10$^4$ yr rapidly oscillates between 10$^{-6}$ $M_\sun$ yr$^{-1}$ and 10$^{-4}$ $M_\sun$ yr$^{-1}$, where its average rate gradually decreases with evolution.
Although the mass accretion rate obtained for CMM3A overlaps the high end of the above range, it is still higher than the average accretion rate.
Note that the dynamical timescale of the outflow might be underestimated, because the SMA observation may miss the older extended outflow components.
Nevertheless, the high accretion rate is robust, even if the dynamical timescale were 10$^4$ yr.
\par
The bolometric luminosity can be estimated from the mass accretion rate as:
\begin{equation}
L_{\rm bol}=\frac{GM_{\rm star}{\dot{M}_{\rm acc}}}{r_{\rm star}}
\end{equation}
where $L_{\rm bol}$ is the bolometric luminosity, $M_{\rm star}$ and $r_{\rm star}$ are the mass and radius of the protostar, respectively, and $\dot{M}_{\rm acc}$ is the mass accretion rate.
Using the protostellar mass of CMM3A (0.1-0.5$M_{\sun}$) and the accretion rate (5 $\times$ 10$^{-5}$ - 4 $\times$ 10$^{-3}$ $M_\sun$ yr$^{-1}$), the average bolometric luminosity is estimated to be as high as 63$\sim$8000~$L_{\sun}$.
Here, we employ the conventional protostellar radius of 2.5 $R_\sun$ \citep[e.g.,][]{Palla_1999, Baraffe_2010}.
This is likely higher than the current bolometric luminosity of $50 \pm 10 \, L_{\sun}$ reported by \citet{Maury_2009}, although the lower end of the estimated range is close to the observed luminosity.
It should be noted that the observed luminosity is not the luminosity of a single protostar, but the total luminosity of CMM3 region.
The luminosity estimated from the accretion rate is regarded as the value over the time from the protostellar birth, which can be different from the current one.
Nevertheless, the difference is large.
\par
This discrepancy on the mass accretion rate could be mitigated in the case of a high accretion rate.
\citet{Hosokawa_2009} pointed out that the protostellar radius can be as high as 100 $R_\sun$ for the accretion rate of 10$^{-3}$ $M_\sun$ yr$^{-1}$.
A larger protostellar radius for a larger accretion rate is also theoretically predicted \citep{Stahler_1986}.
Although the evolution of the protostellar radius is complicated \citep[e.g.,][]{Kuiper_2018}, the above conventional assumption of the protostellar radius of 2.5 $R_\sun$ is likely underestimated for CMM3A.
If it were 100 $R_\sun$, the luminosity is roughly estimated to be  $2-630\,L_\sun$.
\par
The origin of the high accretion rate of CMM3A is not definitively clear but may be related to the binary formation in the environment of NGC2264 cluster.
In this relation, the accretion rate for CMM3B is of particular interest. 
Moreover, we have revealed that the disk/envelope structure is significantly different between CMM3A and CMM3B: the latter harbors more compact disk and more extended envelope than the former. 
This difference also seems to be related to the mechanism of the binary formation. 
Thus, we need the protostellar mass and the accretion rate of CMM3B for further investigations on the origin and the fate of this binary system. 
For this purpose, centimeter-wave line observations, where the dust opacity is significantly lower than the present study case, are essential, as revealed for NGC1333~IRAS4A by \citet{De_Simone_2020}. 
Since NGC2264 CMM3 is relatively close to the Sun and its structure can readily be investigated in detail, it may provide us with important information on the early stage of intermediate-mass protobinary systems. 

\subsection{A Hint of Rotation of the Outflow}
\begin{figure*}[t]
\centering
\includegraphics[clip,width=18cm,page=1,bb = 0 0 1300 600, scale = 1.0]{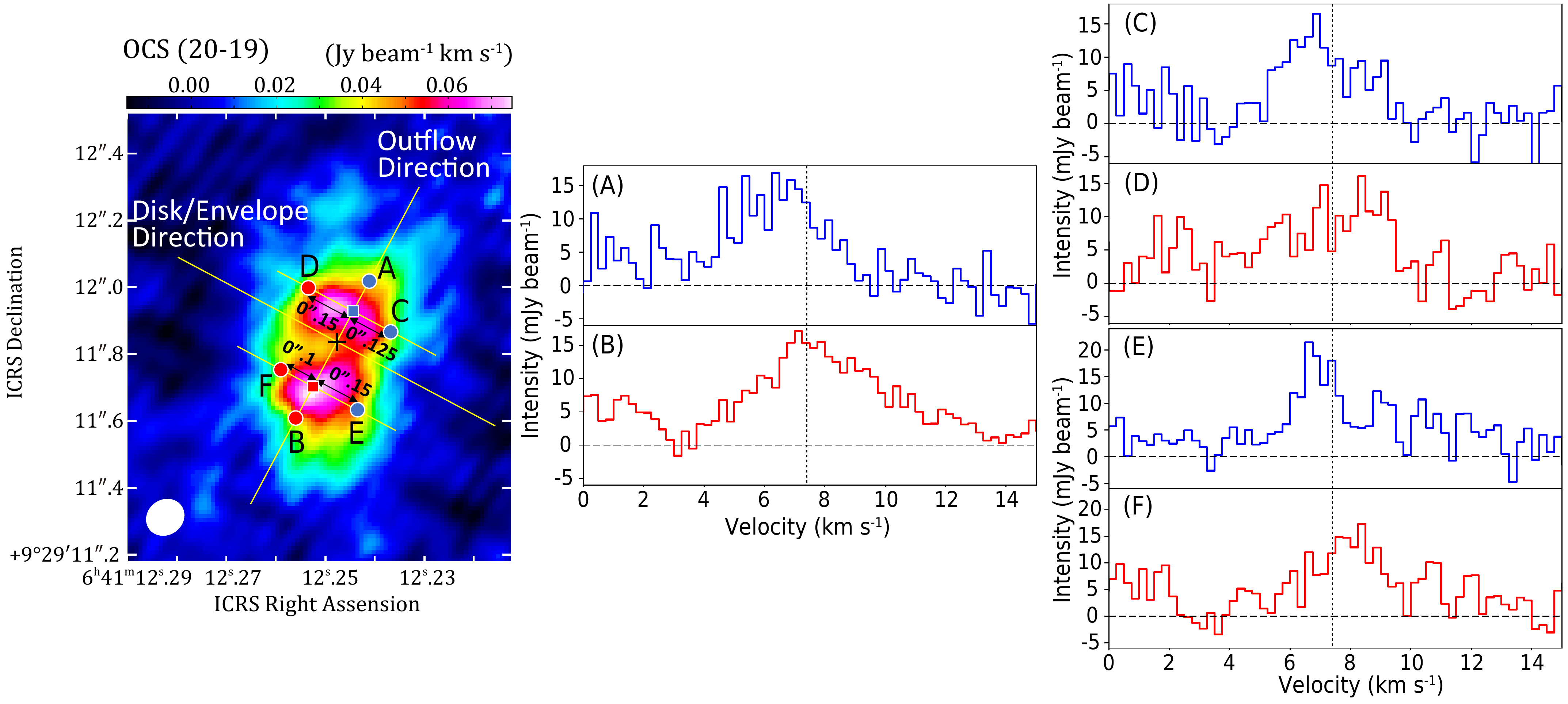}
\caption{The spectrum of OCS ($J=20-19$) emission.
The black cross represents the protostellar position of CMM3A.
The white ellipse represents the beam size.
The vertical and horizontal lines of each panel represent the systemic velocity of CMM3A \citep[7.4km s$^{-1}$:][]{Watanabe_2017} and the ground level (0 Jy beam$^{-1}$), respectively.
}
\label{fig12}
\end{figure*}

The 0th moment map of OCS (Figure~\ref{fig09}) shows that the emission around CMM3A extends along the outflow direction over the disk region traced by the continuum image.
The 1st moment map of OCS (Figure~\ref{fig10}) shows the velocity gradient in the outflow direction.
The blueshifted velocity is dominant in the northwestern part, while the redshifted velocity is dominant in the southeastern part.
Thus, the OCS line likely traces the foot region of the outflow of CMM3A in addition to the disk/envelope system.
In Figure~\ref{fig10}, a velocity gradient across the outflow lobes is marginally visible.
Figure~\ref{fig12} shows the spectral line profiles of OCS toward several positions in the outflow.
The velocity difference is clearly seen between the eastern and western edges of the outflow lobes: the eastern edges are slightly redshifted, while the western edges are slightly blueshifted.
This also suggests a velocity gradient across the outflow lobes. 
This velocity gradient can be interpreted most likely as the rotating motion of the outflow, as revealed in other sources (Orion KL Source I: \citealt{Hirota_2017_na}; NGC1333 IRAS4C: \citealt{Zhang_2018}; L483: \citealt{Oya_2018}).
From Figure~\ref{fig12}, the rotating velocity of the outflow along the line of sight is $\sim$1.0 km s$^{-1}$ at the radius of $\sim$0\farcs15 ($\sim$100 au).
Assuming an inclination angle of 65$^\circ$, the specific angular momentum of the outflow is estimated to be $\sim$6 $\times$ 10$^{-4}$ km s$^{-1}$ pc, which is roughly comparable to the specific angular momentum of the outflow from low-mass protostars \citep[e.g.,][]{Zhang_2018, Oya_2018}.
Here, a caveat is that the velocity gradient across the outflow lobes could reflect the rotating motion of the ambient envelope.
Nevertheless, we think that the outflow rotation case is more reasonable, considering the morphology of the OCS emission extending along the outflow direction.
The 0th moment map of the OCS emission also shows a faint emission in CMM3B, which is distributed along the outflow direction of CMM3B \citep[in the northeast-southwest direction:][]{Watanabe_2017}.
However, the emission line is so faint around CMM3B that the velocity gradient in the disk/envelope direction is not detected in this observation.
\par
For protostars to evolve, the accreting gas needs to lose its angular momentum.
The rotation of the outflow is suggested as a possible candidate of the mechanism of angular momentum extraction \citep[e.g.,][]{Tomisaka_2002, Anderson_2003, Machida_2008, Oya_2018}.
To take the angular momentum away, the outflow needs to have more specific angular momentum (angular momentum divided by its mass) than the accreting gas.
If the observed velocity gradient across the outflow lobes originates from the outflow rotation, the specific angular momentum of the outflow is estimated above to be $\sim$6 $\times$ 10$^{-4}$ km s$^{-1}$ pc, which is comparable to the specific angular momentum of the Keplerian disk at a radius of 35-170~au for the range of the protostellar mass (0.1-0.5 $M_\sun$).
This radius would provide us with rough estimate of the launching point of the outflow \citep{Zhang_2018}.
Such a relatively large launching radius suggests the disk wind caused by magnetocentrifugal effect \citep[e.g.,][]{Konigl_2000, Machida_2008, Machida_2013}.
However, we apparently need higher quality data for the disk rotation and the outflow rotation for further quantitative analyses.

\section{Summary}
We conducted high-angular-resolution observations of NGC 2264 CMM3 with ALMA.
The principal results of this study are summarized as follows.
\par
1.
Continuum images of CMM3A and CMM3B observed at a beam size of 0\farcs1 $\times$ 0\farcs1 ($\sim$70 au) are decomposed to a compact component and an extended component by double 2D-Gaussian fitting.
The compact component of CMM3A seems to trace the disk structure.
The inclination angle of the disk of CMM3A is estimated to be larger than 65$^\circ$.
\par
2.
The spectral index ($\alpha$) between 0.8 mm and 2.0~mm is evaluated from the total flux to be 2.4-2.7 and 2.4-2.6 for CMM3A and CMM3B, respectively.
Considering that CMM3A and CMM3B are young protostars ($\sim$1000 yr), the spectral index suggests high dust opacity of CMM3A and CMM3B, although a possibility of the grain growth cannot be ruled out completely.
Thus, the very sparse line emission in CMM3B likely originates from the dust opacity effect.
\par
3.
The distributions of emission lines of CH$_3$CN and CH$_3$OH in CMM3A are found to have an intensity depression in the disk midplane.
This can be explained by the high optical depths of the dust emission and the line emission.
\par
4.
In the emission lines of CH$_3$CN, CH$_3$OH, and $^{13}$CH$_3$OH, we find the rotation motion of CMM3A.
The protostellar mass of CMM3A is roughly evaluated to be about 0.1-0.5 $M_\sun$ by assuming Keplerian rotation.
The mass accretion rate of CMM3A is estimated to be 5 $\times$ 10$^{-5}$ - 4 $\times$ 10$^{-3}$ $M_\sun$ yr$^{-1}$.
This rate is higher than that of typical low-mass protostars.
\par
5.
We find a hint of the rotating motion of the outflow of CMM3A in the OCS line.
Since the specific angular momentum of the outflow is comparable to the disk, this rotating outflow can contribute to the extraction of the angular momentum from the accreting gas.
\acknowledgements
This study used the ALMA data set ADS/JAO.ALMA\#2018.1.01647.S.
ALMA is a partnership of the European Southern Observatory, the National Science Foundation (USA), the National Institutes of Natural Science (Japan), the National Research Council (Canada), and the NSC and ASIAA (Taiwan), in cooperation with Republic of Chile.
The Joint ALMA Observatory is operated by ESO, the AUI/NRAO, and NAOJ.
The authors acknowledge the ALMA staff for their excellent support.
This study is supported by Grant-in-Aids from Ministry of Education, Culture, Sports, Science, and Technologies of Japan (18H05222, 19H05069, and 19K14753).

\bibliography{reference.bib}{}
\bibliographystyle{aasjournal}

\appendix
\section{Results of 2D-Gaussian fit in the 1.3~mm band}

\begin{figure*}
\centering
\includegraphics[clip,width=18cm,bb = 0 0 1150 1350, scale = 1.0]{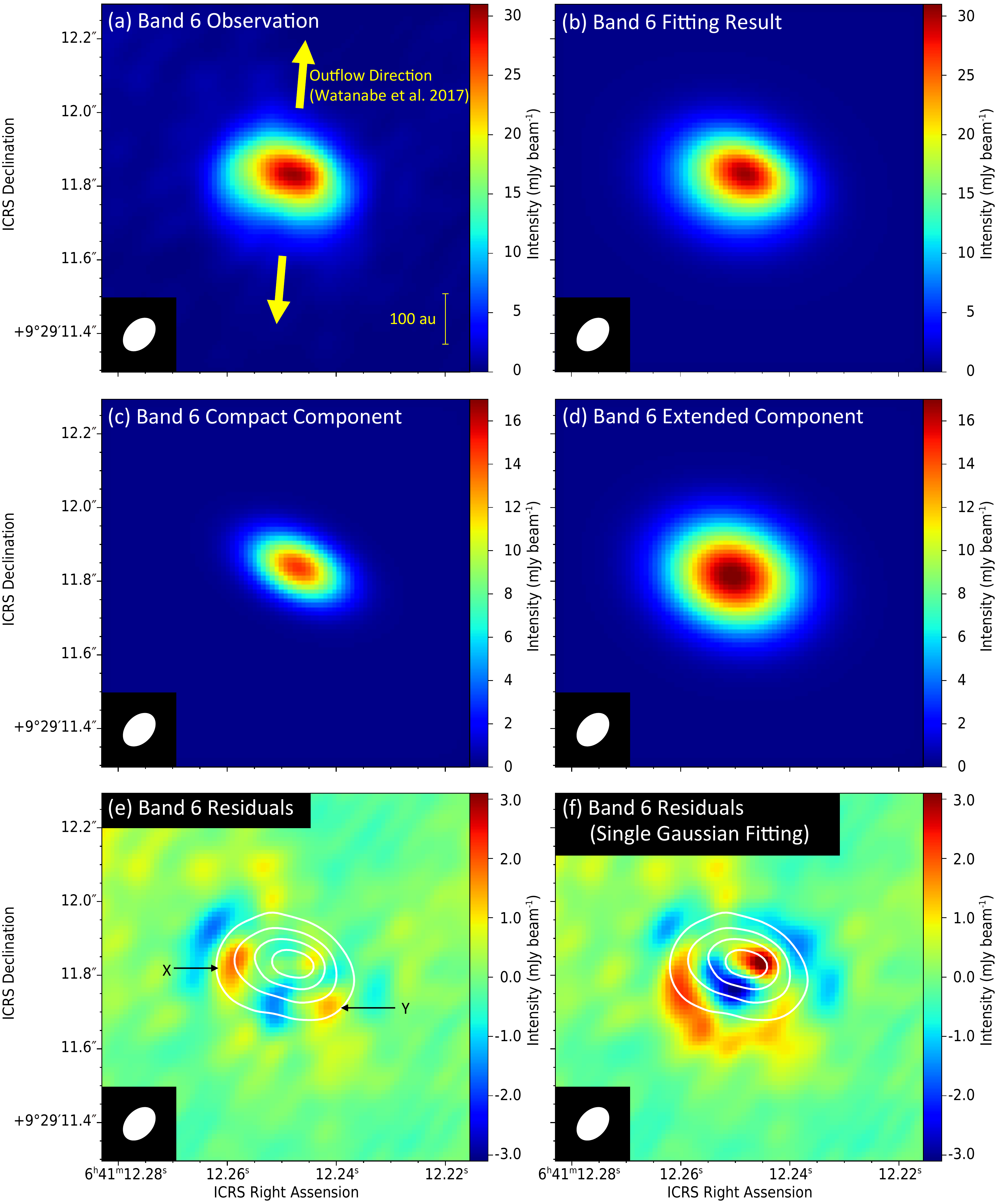}
\caption{(a) The 1.3 mm (Band 6) continuum image of CMM3A.
(b) The result of double 2D-Gaussian fit.
(c), (d) The compact and extended components of the fitted double 2D-Gaussian convolved with the beam.
(e) Residuals of the double 2D-Gaussian fitting.
(f) Residuals of the single 2D-Gaussian fitting for comparison.
The continuum emission at 1.3 mm is displayed in contours with steps of 50$\sigma$ where $\sigma$ is 1.3 $\times$ 10$^{-4}$ Jy~beam$^{-1}$ in (e) and (f).
X and Y denote the position of the systematic residuals also seen in Figure 2.}
\label{fig02}
\end{figure*}

\begin{figure*}
\centering
\includegraphics[clip,width=18cm,bb = 0 0 1150 1350, scale = 1.0]{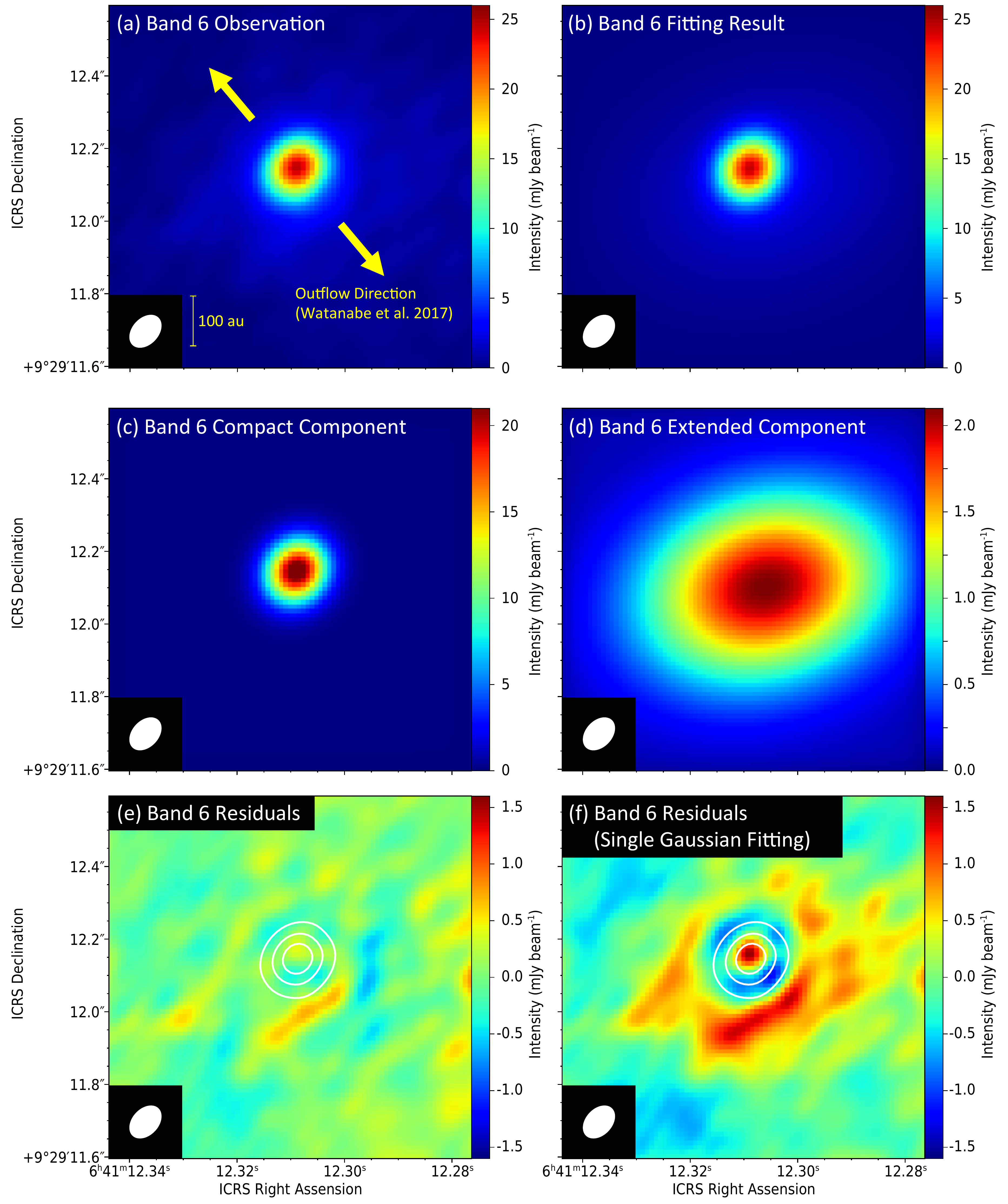}
\caption{(a) The 1.3 mm (Band 6) continuum image of CMM3B.
(b) The result of the double 2D-Gaussian fitting.
(c), (d) The compact and extended components of the fitted double 2D-Gaussian convolved with the beam.
Note that the color scale is different between (c) and (d).
(e) Residuals of the double 2D-Gaussian fitting.
(f) Residuals of the single 2D-Gaussian fitting for comparison.
The continuum emission at 1.3 mm is displayed in contours with steps of 50$\sigma$ where $\sigma$ is 1.3 $\times$ 10$^{-4}$ Jy~beam$^{-1}$ in (e) and (f).
}
\label{fig04}
\end{figure*}

We present the results of 2D-Gaussian fit toward CMM3A (Figure~\ref{fig02}) and CMM3B (Figure~\ref{fig04}) in the 1.3~mm band.
The single 2D-Gaussian fit for CMM3A and CMM3B reveals residual similar to the corresponding case for the 2.0~mm data (Section 3.2), as shown in Figure~\ref{fig02}(f) and Figure~\ref{fig04}(f), respectively.
Then, we conduct the double 2D-Gaussian fit as in the case of the 2.0~mm data (Figure~\ref{fig02}(a-d) and Figure~\ref{fig04}(a-d) for CMM3A and CMM3B, respectively.  The derived parameter are shown in Table~2.  The systematic residuals are almost eliminated (Figure~\ref{fig02}(e) and Figure~\ref{fig04}(e) for CMM3A and CMM3B, respectively).

\clearpage

\end{document}